# Dirac Equation and Planck-Scale Quantities

Rainer Collier

Institute of Theoretical Physics, Friedrich-Schiller-Universität Jena

Max-Wien-Platz 1, 07743 Jena, Germany

E-Mail: rainer@dr-collier.de

**Abstract.** This paper investigate whether quantities with Planck dimensions occur already in the common quantum theory with local Lorentz symmetry . As the quantities $L_{Pl}$ and $M_{Pl}$ involve the Planck constant $\hbar$, the velocity of light $c$ and the gravitational constant $G$ ($M_{Pl} = \sqrt{\hbar c/G}$, $L_{Pl} = GM_{Pl}/c^2$), the relativistic Dirac equation ($\hbar, c$) in the Newtonian gravitational potential ($G$) is considered as a test theory. The evaluation of the break-off condition for the power series of the radial energy eigenfunctions of this purely gravitational atom leads to exact terms for the energy eigenvalues $E_n$ for various special cases of the quantum numbers $N$, $k$ and $n = N + |k|$. It turns out that a useful model of an atom, based solely on Newtonian gravitational forces, can result if, inter alia, the test mass $m_0$ in the gravitational field of the mass $M$ satisfies the condition $m_0 \leq M_{Pl}$.

## 1 Introduction

In the last few decades, the terms of Planck length and Planck energy have frequently turned up in the literature on quantum gravity. It is believed that, in spatial ranges of the linear extension $L \sim L_{Pl} = \sqrt{G\hbar/c^3} = GM_{Pl}/c^2$, the space-time manifold is of a grainy or at least no longer continuous structure. Various approaches to a theory of quantum gravity, such as the string theory, loop quantum gravity, non-commutative geometries, approaches to deformed Lorentz and Poincaré algebras, doubly special relativity (DSR) and diverse generalized uncertainty principles (GUP), therefore, attempt to incorporate into the structure of the theory an elementary length that is independent of the observer and proportional to the Planck length. For an overview of this subject area, see [1], [2] and [3].

The present treatise will look into the question in what way physical quantities with Planck dimensions can occur already in the common quantum theory with local Lorentz symmetry. Since in the Planck quantities $L_{Pl}$ and $M_{Pl}$ there occur Planck's constant $\hbar$, the velocity of light $c$ and Newton's gravitational constant $G$, the relativistic Dirac equation (with $\hbar$ and $c$) in the Newtonian gravitational potential (with $G$) can be considered as a suitable example.

## 2 The Dirac equation with an electromagnetic field

For a better overview of the symbols and terms used, let us first look at the Dirac equation with an electromagnetic field in the flat Minkowski space (with the metric signature and the definition of the four-quantities following the textbook [15]),

$$\left[\underline{\gamma}^k p_k - m_0 c\right]\psi = 0 \quad , \quad p_k = P_k - \frac{q}{c} A_k \quad , \quad P_k = i\hbar \partial_k \quad . \tag{2.1}$$

Here, $\underline{\gamma}^k$ denotes the Dirac matrices, $p^k = (E/c, \vec{p})$ the kinetic four-momentum, $P^k$ the canonical four-momentum, $A^k = (\varphi, \vec{a})$ the four-potential, $x^k = (ct, \vec{r})$ the four-coordinates and $\partial_k = \partial/\partial x^k = (\partial/\partial(ct), \vec{\nabla})$ the four-gradient. The $\underline{\gamma}^k$ matrices satisfy the common commutation rules

$$\underline{\gamma}^k \underline{\gamma}^l + \underline{\gamma}^l \underline{\gamma}^k = 2\eta^{kl} \quad , \tag{2.2}$$

with $\eta^{kl} = diag\,(1,-1,-1,-1)$ denoting the Minkowski tensor. Let the Latin indices $k,l,....$ run from 0 to 3, but the first letters $a,b,....$ from 1 to 3 only.

### 2.1 The Dirac equation in the Coulomb potential

If we have a positive electric charge $Q = Z_e e_0$ situated fixed at the coordinate origin, and a negative electric charge $q = -e_0$ at a distance of $r = |\vec{r}|$, the electric interaction potential $U_{el}$ (in the Gaussian cgs system) has the form

$$U_{el} = q\varphi = \frac{qQ}{r} = -\frac{Z_e e_0^2}{r} \quad , \quad Z_e = \left|\frac{Q}{e_0}\right| \quad , \tag{2.3}$$

where $e_0 = |e|$ is the absolute value of the electric elementary charge. With the corresponding four-potential

$$A^k = (\varphi, \vec{a}) = (\varphi, 0, 0, 0) \quad , \quad \varphi = \frac{Q}{r} \quad , \tag{2.4}$$

the special-relativistic Dirac equation then reads

$$[\underline{\gamma}^k (P_k - \frac{q}{c} A_k) - m_0 c]\psi = 0 \tag{2.5}$$

and, with (3+1) splitting,

$$[\underline{\gamma}^0 (P_0 - \frac{q}{c} A_0) + \underline{\gamma}^a P_a - m_0 c]\psi = 0 \quad . \tag{2.6}$$

With the well-known definitions $\underline{\gamma}^0 = \beta$ and $\beta\underline{\gamma}^a = \alpha^a$, and with definitions (2.1) and (2.2) being considered, multiplication by $c\beta$ (in 2.6) yields

$$[-c\alpha^a P_a + c P_0 - q\varphi - \beta m_0 c^2]\psi = 0 \quad , \tag{2.7}$$

$$[c\vec{\alpha}\cdot\vec{p} + E - q\varphi - \beta m_0 c^2]\psi = 0 \quad . \tag{2.8}$$

With $E\psi \to i\hbar\dot\psi$ and $\vec{p}\psi \to (\hbar/i)\vec\nabla\psi$, the Dirac equation shows the Hamiltonian structure

$$i\hbar\dot\psi = [c\vec\alpha\cdot\vec p + q\varphi + \beta m_0 c^2]\psi \equiv H\psi \quad . \tag{2.9}$$

With the ansatz

$$\psi = \chi(x^a)\exp(Et/i\hbar) \tag{2.10}$$

there follows the time-independent equation of the energy eigenvalues ($E_0 = m_0 c^2$),

$$E\chi = [c\vec\alpha\cdot\vec p + q\varphi + \beta E_0]\chi \quad . \tag{2.11}$$

With the aid of the Dirac matrices ($\vec\sigma$ Pauli matrices)

$$\underline{\gamma}^0 = \begin{pmatrix} \mathbf{1} & 0 \\ 0 & -\mathbf{1} \end{pmatrix} \quad , \quad \underline{\vec\gamma} = \begin{pmatrix} 0 & \vec\sigma \\ -\vec\sigma & 0 \end{pmatrix} \quad , \quad \vec\alpha = \underline{\gamma}^0\underline{\vec\gamma} = \begin{pmatrix} 0 & \vec\sigma \\ \vec\sigma & 0 \end{pmatrix} \quad , \quad \beta = \underline{\gamma}^0 \tag{2.12}$$

and, by decomposing the bispinor $\chi = (\chi^G, \chi^K)^T$, we obtain from (2.11) the set of equations for determining the energy eigenvalues $E$,

$$c\vec\sigma\cdot\vec p\,\chi^K + (U_{el} - E + E_0)\chi^G = 0 \quad , \tag{2.13}$$

$$c\vec\sigma\cdot\vec p\,\chi^G + (U_{el} - E - E_0)\chi^K = 0 \quad . \tag{2.14}$$

With $U_{el}$ from (2.3), $\vec p = -i\hbar\vec\nabla$ and the usual power series for the spinors $\chi^G, \chi^K$, there results the known break-off condition of the radial energy eigenfunctions of the electric atom

$$\sqrt{E_0^2 - E^2}\left(N + \sqrt{k^2 - (Z_e\alpha_e)^2}\right) = (Z_e\alpha_e)E \quad , \tag{2.15}$$

with which the boundary conditions for $\chi^G$ and $\chi^K$ are met. By solving the equation for $E$, we get the discrete energy eigenvalues $E_n$ (Sommerfeld's fine structure formula),

$$E_n \to E_{N,k} = \frac{1}{\sqrt{1 + \frac{(Z_e\alpha_e)^2}{(a_{N,k})^2}}} \quad , \quad a_{N,k} = N + \sqrt{k^2 - (Z_e\alpha_e)^2} \quad , \tag{2.16}$$

$$N = 0, 1, 2, 3, \ldots \quad , \quad k = \pm 1, \pm 2, \pm 3, \ldots \quad , \quad \alpha_e = \frac{e_0^2}{\hbar c} \quad . \tag{2.17}$$

Here, $N$ is the radial quantum number, $k = \pm(j + 1/2)$ with $j$ the total angular momentum quantum number, $n = N + |k|$ the principal quantum number, and $\alpha_e$ the fine structure constant.

## 2.2 Evaluation of the electric break-off condition

The lowest energy level of a quantum system is known to be its ground state. Here, it is the one for $N = 0$ and $k = 1$, i.e. for $n = 1$,

$$E_1 = \frac{E_0}{\sqrt{1 + \left(\frac{Z_e \alpha_e}{a_{0,1}}\right)^2}} = E_0 \sqrt{1 - (Z_e \alpha_e)^2} = E_0 \sqrt{1 - Z_e^2 \left(\frac{e_0^2}{\hbar c}\right)} \quad . \tag{2.18}$$

Hence, the binding and ionization energies relate as

$$E_{bind} = E_1 - E_0 = m_0 c^2 \left(\sqrt{1 - (Z_e \alpha_e)^2} - 1\right) = -E_{ioniz} \quad . \tag{2.19}$$

Note that, with an atomic number $Z_e = 1/\alpha_e \approx 137$, the ground-state energy $E_1$ vanishes, and the ionization energy just equals the energy for creating the electron mass $m_0$. Therefore, according to this simplified atomic model (assuming point masses, neglecting the movement of the atomic nucleus, no radiation corrections,…), atoms with atomic numbers $Z_e > 137$ should not exist.

In preparation for the gravitational case, let us directly evaluate the break-off condition (2.15) for a number of special cases without using the equation form solved in terms of energy $E$ (2.16). By substitution of $x = E/E_0$, eq. (2.15) takes the form

$$\sqrt{1 - x^2} \left(\sqrt{k^2 - (Z_e \alpha_e)^2} + N\right) = (Z_e \alpha_e) \cdot x \quad . \tag{2.20}$$

Now we exactly solve (2.20) for the following cases:

<u>$N \to \infty$</u>

In this case of extreme excitation, we get

$$x \to 1 \quad , \quad E = E_0 = m_0 c^2 \quad . \tag{2.21}$$

<u>$N = 0$</u>

These are radial „ground states" with any angular momenta,

$$\sqrt{1-x^2} \cdot \sqrt{k^2 - (Z_e \alpha_e)^2} = (Z_e \alpha_e) \cdot x \quad , \tag{2.22}$$

$$x = \frac{\sqrt{k^2 - (Z_e \alpha_e)^2}}{|k|} \quad , \quad E = E_0 \sqrt{1 - \frac{(Z_e \alpha_e)^2}{k^2}} \quad . \tag{2.23}$$

$\underline{N = |k|}$

Equal radial and angular momentum excitation,

$$\sqrt{1-x^2} \cdot \left(\sqrt{k^2 - (Z_e \alpha_e)^2} + |k|\right) = (Z_e \alpha_e) \cdot x \quad , \tag{2.24}$$

$$x = \frac{1}{\sqrt{2|k|}} \sqrt{|k| + \sqrt{k^2 - (Z_e \alpha_e)^2}} \quad , \quad E = \frac{E_0}{\sqrt{2}} \sqrt{1 + \sqrt{1 - \frac{(Z_e \alpha_e)^2}{k^2}}} \quad . \tag{2.25}$$

$\underline{Z_e \alpha_e \ll |k|}$

Here, we expand eq. (2.20) according to sufficiently small $\alpha_e$, continuing the series up to the order that is proportional to $(\alpha_e)^4$:

$$x = 1 - \frac{1}{2} \frac{(Z_e \alpha_e)^2}{n^2} - \frac{1}{8} \frac{4N + |k|}{(N+|k|)^4 |k|} (Z_e \alpha_e)^4 + O\left((Z_e \alpha_e)^6\right) \quad , \quad N + |k| = n \quad , \tag{2.26}$$

$$E = E_0 \left[1 - \frac{1}{2} \frac{(Z_e \alpha_e)^2}{n^2} + \left(\frac{3}{8} - \frac{n}{2|k|}\right) \frac{(Z_e \alpha_e)^4}{n^4} + O\left((Z_e \alpha_e)^6\right)\right] \quad . \tag{2.27}$$

$\underline{(Z_e \alpha_e) \lesssim |k|}$

We expand eq. (2.20) according to $(Z \alpha_e)$ values which are sufficiently close to the angular momentum values $|k|$,

$$x = \frac{E}{E_0} = \frac{N}{\sqrt{N^2 + k^2}} + \sqrt{2|k|} \frac{k^2}{\left(\sqrt{N^2 + k^2}\right)^3} \sqrt{|k| - (Z_e \alpha_e)} + O\left(\left(\sqrt{|k| - (Z_e \alpha_e)}\right)^2\right) \quad . \tag{2.28}$$

$\underline{(Z_e \alpha_e) = |k|}$

From (2.28) we can read the solution for this limit case,

$$x = \frac{E}{E_0} = \frac{N}{\sqrt{N^2 + k^2}} \quad . \tag{2.29}$$

Note that all these special formulae can also be obtained by substitution in the exact overall solution for $E$ (formula 2.16).

## 3 The Dirac equation in the Newtonian gravitational potential (Version I)

Consider a purely gravitationally bound quantum system consisting of a central mass $M$ and a test mass $m_0$ (with spin $\hbar/2$) at a distance of $r$. Here again, let the mass $M$ that creates the gravitational field be firmly fixed at the coordinate origin, so that Newton's interaction potential has the form

$$U_{gr} = m_0 \phi = -G \frac{m_0 M}{r} \quad . \tag{3.1}$$

The simplest way to obtain a Dirac equation in the Newtonian gravitational field is via the four-momentum relation generalized for curved spaces having the metric $g_{\mu\nu}$,

$$g_{\mu\nu} p^\mu p^\nu = (m_0 c)^2 \quad , \quad p^\mu = m_0 u^\mu \quad . \tag{3.2}$$

Here, $u^\mu = dx^\mu/d\tau$ denotes the four-velocity and $p^\mu$ the four-momentum of a particle having the rest mass $m_0$. Let the Greek indices $\mu, \nu, ....$ run from 0 to 3, whereas the first letters $\alpha, \beta, ....$ run from 1 to 3 only.

To find out which component of the momentum $p^\mu$ can be identified with the energy constant $E$, let us examine the general-relativistic equation of motion (the geodesic equation) of a test particle of the mass $m_0$ in the gravitational field $g_{\mu\nu}$ of the mass $M$,

$$\frac{Dp_\mu}{D\tau} = p_{\mu;\lambda} u^\lambda = \frac{dp_\mu}{d\tau} - \Gamma^\kappa_{\mu\lambda} p_\kappa u^\lambda = 0 \quad , \tag{3.3}$$

$$\frac{dp_\mu}{d\tau} = m_0 \, \Gamma^\kappa_{\mu\lambda} u_\kappa u^\lambda = m_0 \, \Gamma_{\kappa,\mu\lambda} u^\kappa u^\lambda \quad , \tag{3.4}$$

$$\frac{dp_\mu}{d\tau} = \frac{m_0}{2} g_{\kappa\lambda|\mu} \, u^\kappa u^\lambda. \tag{3.5}$$

Now, to make the equation of motion (3.5) congruent with Newton's equation of motion, let the metric $g_{\mu\nu}$ have the following form:

$$g_{00} = 1 + \frac{2\phi}{c^2} \quad , \quad g_{\alpha\beta} = -\delta_{\alpha\beta} \quad , \tag{3.6}$$

with $\phi = \phi(x^\alpha)$ being a function of the spatial coordinates only. Since all potentials $g_{\kappa\lambda}$ are independent of the time coordinate $x^0$, then $dp_0/d\tau = 0$, with $p_0 = E/c = const.$ being a conserved quantity. Let us identify this quantity with the total energy $E$ of the motion.

Let it be emphasized that we are treating a simple model system, in which only the time-time component $g_{00}$ of the metric tensor $g_{\mu\nu}$ is non-zero, whereas its spatial component $g_{\alpha\beta} = -\delta_{\alpha\beta}$ belongs to a flat position space. Thus, we do not deal with the case of the Dirac equation in the weak Schwarzschild field, which is known to have a weak position space curvature. With this metric, i.e., $g_{\mu\nu} = diag\,(g_{00}, -1, -1, -1)$ and $p_\alpha = -p^\alpha$, (3.5) becomes a Newtonian equation of motion,

$$\frac{dp^\alpha}{d\tau} = -\frac{m_0}{2} g_{00|\alpha} u^0 u^0 = -m_0 \phi_{|\alpha} \lambda^2 \quad , \quad \lambda d\tau = dt \quad , \tag{3.7}$$

$$\frac{d\vec{p}}{dt} = \vec{F} \quad , \quad \vec{p} = m\vec{v} \quad \vec{F} = -m\vec{\nabla}\phi \quad , \quad m = \lambda m_0 \; . \tag{3.8}$$

Note that (3.8) is a relativistically generalized Newtonian equation of motion, as $\vec{p}$ is the relativistic momentum and $\vec{F}$ is a relativistic generalization of the Newtonian gravitational force. Therefore, the motion of the mass $m_0$ in the gravitational field of the mass $M$ need not be restricted to, e.g., non-relativistic velocities.

What, now, is the structure of the conservation law $p_0 = E/c = const.$? Proceeding from (3.2) and referring to the metric (3.6), we get

$$g^{00} p_0^{\,2} + g^{\alpha\beta} p_\alpha p_\beta = g^{00}\left(\frac{E}{c}\right)^2 - \delta^{\alpha\beta} p_\alpha p_\beta = (m_0 c)^2 \quad , \tag{3.9}$$

which is the desired relativistic energy equation for the motion of a mass point in the Newtonian gravitational field,

$$\sqrt{g^{00}}\, E \approx \left(1 - \frac{\phi}{c^2}\right) E = \sqrt{c^2 \vec{p}^{\,2} + (m_0 c^2)^2} = E_{kin} \quad , \tag{3.10}$$

$$E \approx E_{kin} + m\,\phi \quad , \quad m = \frac{E}{c^2} = m_0 + \left(\frac{m_0}{2} v^2 + \cdots + m_0 \phi + \cdots\right)\frac{1}{c^2} \; . \tag{3.11}$$

Here, $m = E/c^2$ is the heavy mass that corresponds to the total (conserved) kinetic and potential energy of the mass $m_0$ in the gravitational field $\phi$ of the mass $M$. What is interesting here is that the gravitational field of the central mass acts on the entire heavy mass $m = E/c^2$ of the test particle.

Squaring in eq. (3.10) immediately leads to

$$g^{00} E^2 - c^2 \vec{p}^{\,2} = (m_0 c^2)^2 \; . \tag{3.12}$$

Using the Dirac matrices (2.2) in the known representation (2.12), we change over to the operator form of the equation, applied to a bispinor $\chi$,

$$\left[\underline{\gamma}^0 \sqrt{g^{00}}\, E + \underline{\gamma}^a c p_a - m_0 c^2\right]\chi = 0 \tag{3.13}$$

Multiplication by $\underline{\gamma}^0 = \beta$, equating $\beta\underline{\gamma}^a = \alpha^a$ and approximating $\sqrt{g^{00}} \approx 1 + \phi/c^2$ yields

$$E\chi = [c\vec{\alpha}\cdot\vec{p} + m\phi + \beta E_0]\chi \quad . \tag{3.14}$$

By splitting the bispinor $\chi(x^\alpha) = (\chi^G, \chi^K)^{\mathrm{T}}$, we obtain the coupled differential equation system

$$c\vec{\sigma}\cdot\vec{p}\chi^K + (U_{gr} - E + E_0)\chi^G = 0 \quad , \tag{3.15}$$

$$c\vec{\sigma}\cdot\vec{p}\chi^G + (U_{gr} - E - E_0)\chi^K = 0 \quad , \tag{3.16}$$

with $\vec{p} = (\hbar/i)\vec{\nabla}$, $U_{gr} = m\phi = (E/c^2)\phi$ and $E_0 = m_0 c^2$.

## 4 The Dirac equation in the Newtonian gravitational potential (Version II)

We proceed from the general-relativistic Dirac equation in the metric $g_{\mu\nu}$:

$$(\gamma^\mu p_\mu - m_0 c)\psi = 0 \quad , \tag{4.1}$$

$$p_\mu = P_\mu - \Gamma_\mu \quad , \quad \Gamma_\mu = \frac{1}{4}\gamma^\kappa \gamma_{\kappa;\mu} \quad , \quad P_\mu = i\hbar\partial_\mu \quad , \tag{4.2}$$

$$\gamma^\kappa \gamma^\lambda + \gamma^\lambda \gamma^\kappa = 2g^{\kappa\lambda} \quad , \quad \gamma_{\kappa;\mu} = \gamma_{\kappa|\mu} - \Gamma^\rho_{\kappa\mu}\gamma_\rho \quad , \tag{4.3}$$

where $\gamma_{\kappa;\mu}$ is Riemann's covariant derivative of $\gamma_\kappa$ with the Christoffel symbols $\Gamma^\rho_{\kappa\mu}$. The bispinor affinities $\Gamma_\mu$ cause the bispinorial covariant derivative of the bispin tensors $\gamma_\kappa$ to vanish,

$$\gamma_{\kappa\|\mu} = \gamma_{\kappa;\mu} - [\Gamma_\mu, \gamma_\kappa] = 0 \quad . \tag{4.4}$$

The commutation rules of the $\gamma^\kappa$ matrices in (4.3) can be satisfied by a set of tetrad four-vectors $e_k^\mu$,

$$\gamma^\mu = e_k^\mu \underline{\gamma}^k \quad , \quad e_k^\mu e_l^\nu \eta^{kl} = g^{\mu\nu} \quad , \tag{4.5}$$

where $\underline{\gamma}^k$ are the Dirac matrices (2.2) in the flat tangential space. For the Newtonian gravitational field, let the metric $g_{\mu\nu}$ according to eq. (3.6) be described by the structure

$$g_{00} = 1 + 2\frac{\phi}{c^2} \quad , \quad g^{00} = \frac{1}{g_{00}} \quad , \quad g_{0\alpha} = 0 \quad , \quad g_{\alpha\beta} = -\delta_{\alpha\beta} \quad . \tag{4.6}$$

For this time-independent metric, only two Christoffel symbols are unequal to zero:

$$\Gamma^0_{\alpha 0} = \frac{1}{2} g^{00} g_{00|\alpha} \quad , \quad \Gamma^\alpha_{00} = -\frac{1}{2} g^{\alpha\beta} g_{00|\beta} \quad . \tag{4.7}$$

The tetrads $e^\mu_k$ have the form

$$e^0_k = \sqrt{g^{00}} \delta^0_k \quad , \quad e^\alpha_k = \delta^\alpha_k \quad ; \tag{4.8}$$

and hence, the metric Dirac matrices $\gamma^\mu$ appear as

$$\gamma^0 = e^0_k \underline{\gamma}^k = \sqrt{g^{00}} \underline{\gamma}^0 \quad , \quad \gamma^\alpha = e^\alpha_k \underline{\gamma}^k = \delta^\alpha_k \underline{\gamma}^k \quad . \tag{4.9}$$

Therewith, the bispinor affinity $\Gamma_\mu$ or the quantity $\Gamma = \gamma^\mu \Gamma_\mu$ can be calculated to be

$$\Gamma = \frac{1}{4} \gamma^\mu \gamma^\kappa \gamma_{\kappa;\mu} = \frac{1}{4} \left( \gamma^0 \gamma^\kappa \gamma_{\kappa;0} + \gamma^\alpha \gamma^\kappa \gamma_{\kappa;\alpha} \right) \quad . \tag{4.10}$$

The second addend vanishes, leaving

$$\Gamma = \frac{1}{4} \gamma^0 g_{00|\alpha} \gamma^0 \gamma^\alpha = \frac{1}{4} g^{00} g_{00|\alpha} \gamma^\alpha \quad . \tag{4.11}$$

Because of $(-g) = -g_{00} g_{11} g_{22} g_{33} = g_{00} = 1 + 2\phi/c^2$ and $g_{00|0} = 0$, we can also write

$$\Gamma = \frac{1}{2} \left( \ln \sqrt{-g} \right)_{|\mu} \gamma^\mu \quad , \quad \Gamma_\mu = \frac{1}{2} \left( \ln \sqrt{-g} \right)_{|\mu} \quad , \quad . \tag{4.12}$$

Therefore, the Dirac equation (4.1) in the Newton potential $\phi$ reads

$$\gamma^\mu \left[ P_\mu - \frac{1}{2} \left( \ln \sqrt{-g} \right)_{|\mu} - m_0 c \right] \psi = 0 \quad . \tag{4.13}$$

By means of a unitary equivalence transformation of the form

$$\gamma^\mu \to C \gamma^\mu C^{-1} \quad , \quad \psi \to C \psi \quad , \quad C = \sqrt[4]{-g} \begin{pmatrix} 1 & 0 \\ 0 & 1 \end{pmatrix} \tag{4.14}$$

we cause the $\ln \sqrt{-g}$ term to vanish. This results in the simple equation

$$(\gamma^\mu P_\mu - m_0 c) \psi = 0 \quad , \quad P_\mu = i\hbar \partial_\mu \quad . \tag{4.15}$$

It is known, by the way, that suitable equivalence transformations can make the bispinor affinity $\Gamma_\mu$ vanish in all orthogonal metrics (even in almost all time-orthogonal metrics) [4]. The role of bispinor affinities in connection with gravito-magnetic effects is investigated in detail in [5], e.g.

Because of the ansatz $\psi = \chi(x^\alpha)\exp(E x^0/i\hbar c)$ in eq. (4.15), there follows first

$$\left[\gamma^0 \frac{E}{c} - \frac{\hbar}{i}\gamma^\alpha \partial_\alpha - m_0 c\right]\chi = 0 \quad . \tag{4.16}$$

With the aid of the $\gamma^\mu$ matrices from eq. (4.5), $\underline{\gamma}^0\underline{\gamma}^a = \alpha^a$ and multiplication by $c\underline{\gamma}^0 = c\beta$, we get with (4.16)

$$\left[\sqrt{g_{00}}\, E - \frac{\hbar}{i} c\alpha^a \partial_a - \beta m_0 c^2\right]\chi = 0 \quad . \tag{4.17}$$

Using $\sqrt{g^{00}} \approx 1 - \phi/c^2$ and substituting $\vec{p} = (\hbar/i)\vec{\nabla}$ again, we get now

$$E\chi = [c\vec{\alpha}\cdot\vec{p} + m\phi + \beta E_0]\chi \quad , \quad m = \frac{E}{c^2} \quad , \quad E_0 = m_0 c^2 \quad . \tag{4.18}$$

This is the quantum equation of the gravitational atom, found already in eq. (3.14) by simple Dirac root extraction. Again we can see from the potential term $m\phi$ that the gravitational field of the central mass $M$ acts on the entire heavy mass $m = E/c^2$ of the test particle rather than on its rest mass $m_0$ only.

A selection of studies on the Dirac equation in gravitational fields, especially in the Schwarzschild field, will be found in references [6] to [14].

## 5 The energy eigenvalues of the gravitational Dirac equation

The stationary energy levels of our gravity atom can be determined from the decisive set of equations (3.14) or (4.18). This is very easy now, because the set of equations (4.18) formally matches that of the electric atom (2.11), provided only that the electric potential $U_{el}$ is suitably converted into the gravitational potential $U_{gr}$. We have

$$U_{el} = q\varphi = \frac{qQ}{r} = -\frac{e_0(Z_e e_0)}{r} = -\frac{\hbar c}{r}Z_e \frac{e_0^2}{\hbar c} = -\frac{\hbar c}{r}Z_e \alpha_e \quad , \tag{5.1}$$

$$U_{gr} = m\phi = -\frac{GmM}{r} = -\left(\frac{m}{m_0}\right)\frac{Gm_0 M}{r} = -\frac{\hbar c}{r}Z_g \frac{Gm_0 M}{\hbar c} = -\frac{\hbar c}{r}Z_g \alpha_g \quad , \tag{5.2}$$

with the electric and gravitational constants being defined as follows:

$$Z_e = \left|\frac{Q}{e_0}\right| \quad , \quad \alpha_e = \frac{e_0^2}{\hbar c} \quad , \tag{5.3}$$

$$Z_g = \frac{m}{m_0} = \frac{E}{E_0} \quad , \quad \alpha_g = \frac{Gm_0 M}{\hbar c} = \frac{m_0 M}{M_{Pl}^2} \quad , \quad M_{Pl} = \sqrt{\frac{\hbar c}{G}} \quad . \tag{5.4}$$

In the break-off condition (2.15) of the electric atom, therefore, we simply replace

$$Z_e \cdot \alpha_e \to Z_g \cdot \alpha_g \tag{5.5}$$

and re-evaluate the resulting break-off condition of the Dirac equation of the gravitational atom with regard to the energy constant $E$ (which is now also contained in $Z_g$),

$$\sqrt{E_0^2 - E^2} \left( N + \sqrt{k^2 - (Z_g \alpha_g)^2} \right) = (Z_g \alpha_g) E \quad , \tag{5.6}$$

$$\sqrt{E_0^2 - E^2} \left( N + \sqrt{k^2 - \left(\frac{E}{E_0}\right)^2 \alpha_g^2} \right) = \left(\frac{E}{E_0}\right) E \, \alpha_g \quad . \tag{5.7}$$

However, the ranges of values of the quantum numbers $N, k$ and $n, j$ are identical to those of the electric case (compare eq. (2.17)).

Such a procedure is possible because, with regard to $r$-dependence and singular behaviour, the set of equations (4.18) has the same mathematical structure as the set (2.11).

**5.1 The ground state of the gravitational atom**

To evaluate eq. (5.6) we equate $E/E_0 = x$ and get

$$\sqrt{1 - x^2} \left( N + \sqrt{k^2 - x^2 \alpha_g^2} \right) = x^2 \alpha_g \quad . \tag{5.8}$$

If we were going to solve this equation, we would encounter a cubic equation in $x^2$, the solution of which by Cardano's formulae would be possible, but hardly appropriate for special statements. Let us rather evaluate eq. (5.7) directly and compare the results with the corresponding cases of the electric atom.

The resulting ground state of the gravitational atom, with $N = 0$ and $k = 1$, i.e., $n = 1$, is

$$E_1 = \frac{E_0}{\sqrt{1 + \alpha_g^2}} = E_0 \frac{M_{Pl}}{\sqrt{M_{Pl}^2 + m_0 M}} \quad . \tag{5.9}$$

Hence, the binding and ionization energies relate as

$$E_{bind} = E_1 - E_0 = m_0 c \left( \frac{1}{\sqrt{1 + \alpha_g^2}} - 1 \right) = -E_{ioniz} \quad . \tag{5.10}$$

It is obvious that the binding energy cannot exceed the amount of $m_0 c^2$, not even for immense values of $\alpha_g$. In contrast to the ground level of the electric atom (2.19), no particular size limit can be seen here for the masses $m_0$ and $M$ involved in the interaction. If we look at the break-off relation (5.8), though, at least the conditions

$$x^2 \alpha_g^2 \leq k^2 \quad , \quad x^2 \leq 1 \tag{5.11}$$

must be kept, i.e., for $|k|=1$,

$$\alpha_g \leq 1 \quad , \quad m_0 M \leq M_{Pl}^2 \quad . \tag{5.12}$$

This requirement for the masses $m_0$ and $M$ readily makes sense physically: The inequality (5.12) can be converted as follows,

$$\alpha_g = \frac{m_0 M}{M_{Pl}^2} = \frac{G m_0 M}{\hbar c} = \frac{GM}{c^2} \cdot \frac{m_0 c}{\hbar} = \frac{r_g}{\lambda_c} \leq 1 \quad , \quad r_g = \frac{GM}{c^2} \quad , \quad \lambda_c = \frac{\hbar}{m_0 c} \quad . \tag{5.13}$$

The condition $\alpha_g \leq 1$ conveys that this gravitational atomic theory is valid only as long as the Compton wavelength $\lambda_c$ of the test mass $m_0$ is always greater than the gravitational length $r_g$ of the field-generating mass $M$ ($r_s = 2 r_g$ is the Schwarzschild radius of $M$). Thus, our gravitational atomic model indicates its own range of validity!

**5.2 Evaluation of the gravitational break-off condition**

Let us now evaluate certain special cases of the break-off condition (5.8), analogously to the procedure with the electric atom in chapter 2. For the following cases, let us solve the condition (5.8) exactly,

$\underline{N \rightarrow \infty}$

In case of extreme excitation, we get

$$N \rightarrow \infty \quad , \quad x \rightarrow 1 \quad , \quad E \rightarrow E_0 = m_0 c^2 \quad . \tag{5.14}$$

$\underline{N = 0}$

Radial ground states with any angular momenta

$$\sqrt{1-x^2} \cdot \sqrt{k^2 - x^2 \alpha_g^2} = \alpha_g x^2 \quad , \tag{5.15}$$

$$x = \frac{|k|}{\sqrt{k^2 + \alpha_g^2}} \quad , \quad E = \frac{E_0}{\sqrt{1 + \frac{\alpha_g^2}{k^2}}} \quad . \tag{5.16}$$

$\underline{N = |k|}$

Simultaneously increasing radial and angular momentum excitation

$$\sqrt{1-x^2}\left(\sqrt{k^2 - x^2\alpha_g^2} + |k|\right) = \alpha_g x^2 \quad , \tag{5.17}$$

$$x = \frac{1}{2|k|}\sqrt{(2k)^2 - \alpha_g^2} \quad , \quad E = E_0\sqrt{1 - \frac{\alpha_g^2}{(2k)^2}} \quad . \tag{5.18}$$

$\underline{\alpha_g \ll |k|}$

Here, let eq. (5.8) be expanded according to a small $\alpha_g$:

$$x = 1 - \frac{\alpha_g^2}{2(N+|k|)^2} - \frac{1}{8}\frac{4N - 3|k|}{|k|(N+|k|)^4}\alpha_g^4 + O(\alpha_g^6) \quad , \quad N + |k| = n \quad , \tag{5.19}$$

$$\frac{E}{E_0} = 1 - \frac{1}{2}\frac{\alpha_g^2}{n^2} + \left(\frac{7}{8} - \frac{n}{2|k|}\right)\frac{\alpha_g^4}{n^4} + O(\alpha_g^6) \quad . \tag{5.20}$$

It can be noted that, up to the term that is proportional to $1/n^3$, the energy spectra $(E_{el})_n$ in (2.27) and $(E_{gr})_n$ in (5.19) have the same hydrogenic form. It is only from the term proportional to $1/n^4$ onward that the level $(E_{gr})_n$ is higher than the corresponding level $(E_{el})_n$ by the amount of $(1/2)(\alpha_g/n)^4$ (if we assume that $\alpha_g = \alpha_e$). It is remarkable that this simple model of a Dirac particle in the Newtonian gravity field in the fine structure domain proportional to $\alpha_g^4$ already brings half ($\frac{7}{8}$) of the exact value ($\frac{15}{8}$) (see the investigations about Dirac particles in the Schwarzschild field in [11] and [14]). The remaining share is obviously contributed by taking the position space curvature into account.

$\underline{\alpha_g \lesssim |k|}$

Eq. (5.8) is expanded according to $\alpha_g$ - values which are very close to the angular momentum values $|k|$:

$$x = \frac{E}{E_0} = x_0 + \frac{y_0}{|k|}\left(|k| - \alpha_g\right) + O\left((|k| - \alpha_g)^2\right) \quad , \tag{5.21}$$

$$x_0 = \frac{1}{4}\sqrt{8 - 2A^2 + 2A\sqrt{A^2 + 8}} \quad , \quad y_0 = \frac{2x_0\sqrt{1 - x_0^2}}{4\sqrt{1 - x_0^2} + A} \quad , \quad A = \frac{N}{|k|} \tag{5.22}$$

$$\underline{\alpha_g = |k|}$$

From this, we can read the solution $x = E/E_0 = x_0$ for the limit case $\alpha_g = |k|$. Note that, for the gravitational atom, an expansion is possible only according to powers of $(|k| - \alpha_g)$, whereas for the electric atom (2.28) we have been able to implement an expansion according to powers of $\sqrt{|k| - \alpha_e}$.

## 6 Conclusions

We considered an atom existing due to a purely gravitational interaction between the two neutral masses $M$ (central mass) and $m_0$ (test mass with spin $\hbar/2$). Let this interaction be described merely by Newton's law of gravitation, i.e., be regarded as being of weak gravity ($U_{gr} \ll m_0 c^2$, i.e., $r \gg r_g$). The investigation was aimed to find out whether, in such a quantum system described by the Dirac equation (with the universal constants $\hbar, c$) in the gravitational field (with universal constant G), there are any hints to Planck quantities that are a combination of these three universal constants (such as, e.g., the Planck mass $M_{Pl} = \sqrt{\hbar c / G}$ and Planck length $L_{Pl} = GM_{Pl}/c^2$).

For this purpose, in chapter 3 an energy conservation law was derived from the general-relativistic equation of motion of a test mass point $m_0$ in the time-independent gravitational field of the central mass $M$ and then in the case of a weak Newtonian interaction potential the Dirac root was extracted. In another way in chapter 4, the general-relativistic Dirac equation was evaluated for the weak Newtonian gravitational potential. Both approaches lead to the same gravitational Dirac equation. The quantum equation thus constructed for a gravitational atomic model (4.18) with Newtonian interaction is analogous to the corresponding Dirac equation for the electric atomic model (2.11) with Coulomb interaction. Both quantum equations take the same mathematical form if the following substitutions are made in the electric Dirac equation,

$$Z_e \cdot \alpha_e \longrightarrow Z_g \cdot \alpha_g \quad , \tag{6.1}$$

$$\left\{ Z_e = \left|\frac{Q}{e_0}\right| \quad , \quad \alpha_e = \frac{e_0^2}{\hbar c} \right\} \longrightarrow \left\{ Z_g = \frac{E}{E_0} \quad , \quad \alpha_g = \frac{m_0 M}{M_{Pl}^2} \right\} \quad , \tag{6.2}$$

Therefore, if the substitutions (6.1), (6.2) are made in the break-off condition (2.15) of the radial energy eigenfunctions of the electric atomic model, one unconventional gets to the corresponding break-off condition (5.6) of the gravitational atomic model. Note that, in this gravitational break-off condition, the energy $E$ to be determined also occurs in the constant $Z_g$. In chapter 5, the break-off condition of the gravitational atom was evaluated for several special cases of the radial quantum number $N$ and of the angular momentum quantum number $k$.

A difference from the analogous energy levels of the electric atomic model with Coulomb interaction (compare (2.27) with (5.20)) occurs only in the approximation proportional to $\alpha_g^4$ (at $\alpha_g \ll 1$). This difference is mainly caused by the substitution of the heavy mass $m = E/c^2$ for the rest mass $m_0$.

The difference between the electric and the gravitational atomic model is particularly marked at the ground state level $E_1$, which belongs to the quantum numbers $N = 0$, $|k|=1$, i.e., $n=1$. The exact value in the electric atom is

$$(E_1)_{el} = E_0\sqrt{1-\alpha_e^2} \approx E_0\left(1 - \frac{1}{2}\alpha_e^2 - \frac{1}{8}\alpha_e^4 + \cdots\right) \quad , \quad \alpha_e = \frac{e_0|Q|}{\hbar c} \quad , \tag{6.3}$$

whereas, in the gravitational atom, we get

$$(E_1)_{gr} = E_0\frac{1}{\sqrt{1+\alpha_g^2}} \approx E_0\left(1 - \frac{1}{2}\alpha_g^2 + \frac{3}{8}\alpha_g^4 + \cdots\right) \quad , \quad \alpha_g = \frac{m_0 M}{M_{Pl}^2} \quad . \tag{6.4}$$

Also, the term for $(E_1)_{el}$ in (6.3) shows a limit behaviour. Necessarily,

$$\alpha_e \leq 1 \quad , \quad |Q| \leq \left(\frac{\hbar c}{e_0^2}\right) \cdot e_0 \approx 137 \cdot e_0 \quad . \tag{6.5}$$

An analogous limit, e.g. for the mass $M$, does not result from the ground-state energy $(E_1)_{gr}$ in (6.4). However, it follows from the general gravitational break-off condition (5.8) that

$$x^2 \leq 1 \quad , \quad x^2 \cdot \alpha_g^2 \leq 1 \quad , \quad \alpha_g = \frac{m_0 M}{M_{Pl}^2} \leq 1 \quad , \quad M_{Pl} = \sqrt{\frac{\hbar c}{G}} \quad . \tag{6.6}$$

The condition $\alpha_g \leq 1$ contained therein only indicates that, if $m_0$ is predetermined, the necessary consequence is the choice of $M \leq M_{Pl}^2/m_0$. If, e.g., $m_0 = m_e$ (the electron mass), $m_e = 9.1 \cdot 10^{-31}$ kg and $M_{Pl} = 2.2 \cdot 10^{-8}$ kg, then $M$ may have a mass of up to $M \leq M_{Pl}^2/m_e = 5.3 \cdot 10^{14}$ kg.

If, however, one selects $m_0 \lesssim M_{Pl}$, $M$ must at least also have one Planck mass $M \gtrsim M_{Pl}$. For still greater test masses $m_0 > M_{Pl}$ and, thus, $M < M_{Pl}$, our atomic model overturns: The test mass $m_0$ would be greater than the mass $M$ generating the gravitational field. Such an assumption, although possible according to (6.6), destroyed the key assumptions of the model and will therefore be excluded. Then, however, the inequality (6.6) also describes some kind of limit behaviour: For the existence of a gravitational atom, it will be true only if the test mass $m_0$ and the mass $M$ generating the gravitational field, respectively, are

$$0 < m_0 \leq M_{Pl} \quad , \quad M_{Pl} \leq M < \infty \quad , \tag{6.7}$$

connected by the inequality (6.6).

In the limit case $m_0 = M = M_{Pl}$, there is the gravitational Planck-mass atom. Its Bohr radius, by the way, is exactly one Planck length. For two masses $m_0, M$, the Bohr radius of an gravitational atom (analogously to the electric atom) is defined by

$$a_g = \frac{\lambda_c}{\alpha_g} = \frac{\hbar}{m_0 c} \cdot \frac{M_{Pl}^2}{m_0 M} = \frac{\hbar}{m_0 c} \cdot \frac{\hbar c}{G m_0 M} = \frac{\hbar^2}{m_0^2 (GM)} \quad . \tag{6.8}$$

In the case of a Planck mass atom, because of $m_0 = M = M_{Pl}$, and $\alpha_g = (m_0 M / M_{Pl}^2) = 1$, we have indeed

$$a_g = \lambda_c = \frac{\hbar}{M_{Pl} c} = \sqrt{\frac{G\hbar}{c^3}} = L_{Pl} \quad , \tag{6.9}$$

and the maxima $r_n$ of the radial probability of presence of the test mass $m_0 = M_{Pl}$ lie at about $r_n \sim L_{Pl} n^2$.

Another notation of the inequality (6.6) is interesting as well. If we use the Compton wavelength of the test mass $m_0$ and the gravitational length $r_g = r_s / 2$ of the field-generating mass $M$, (6.6) also reads

$$\alpha_g = \frac{m_0 M}{M_{Pl}^2} \leq 1 \quad , \quad \lambda_c = \frac{\hbar}{m_0 c} \geq \frac{GM}{c^2} = r_g \quad . \tag{6.10}$$

Consequently, this gravitational atomic model is valid only as long as the Compton wavelength $\lambda_c$ of the test mass $m_0$ always remains greater than the gravitational length $r_g = r_s / 2$ of the field-generating mass $M$. This statement follows from the mathematics of the model itself.

The examinations of a gravitational atomic model have shown that Planck quantities such as $L_{Pl}, M_{Pl}$ or $E_{Pl}$, occurring in this special quantum system, do not constitute universal limits of lengths or energies. For an atomic model dealt with on the basis of a strictly relativistic theory, this was to be expected. In case of the atomic model discussed here on the basis of a purely gravitational interaction, however, the inequality (6.6) together with the model conditions (6.7) do point to such Planck scale limits.

# 7 Appendix

Between the difficult explicit solution of the gravitational break-off condition (5.8) and the treatment of special cases in chapter 5.2, there is an interesting approximation version for a very small $\alpha_g \ll 1$, which Barros [8] used in his examination of the hydrogen atom with inner Schwarzschild metric generate by the electric interaction energy $U_{el} = qQ/r$.

In our gravitational break-off condition (5.8)

$$\sqrt{1-x^2}\left(N + \sqrt{k^2 - x^2 \alpha_g^2}\right) = x^2 \alpha_g \quad , \tag{7.1}$$

we can assume that, for $\alpha_g \ll 1$, $x = E/E_0 \approx 1$ and hence (with $|k| \geq 1$) may be put under the root $x^2 \alpha_g^2 \approx \alpha_g^2$. This yields a break-off condition that is slightly corrected compared to (7.1),

$$\sqrt{1-x^2}\left(N + \sqrt{k^2 - \alpha_g^2}\right) = x^2 \alpha_g \quad . \tag{7.2}$$

An equation of that structure can be found in Barros´ paper. It can be solved straightforwardly for $x$ and, thus, for $E$:

$$E_n = E_0 \sqrt{\frac{2}{1 + \sqrt{1 + \frac{4\alpha_g^2}{a_n^2}}}} \quad , \quad a_n = N + \sqrt{k^2 - \alpha_g^2} \quad , \quad n = N + |k| \quad . \tag{7.3}$$

By the form of $a_n$ it can be seen again that the condition $\alpha_g = (m_0 M / M_{Pl}^2) \leq 1$ has to be respected in the gravitational atom as well. Assuming that $(\alpha_g^2 / a_n^2) \ll 1$, one can write for (7.3), according to Barros,

$$E_n = \frac{E_0}{\sqrt{1 + \frac{\alpha_g^2}{a_n^2} - \frac{\alpha_g^4}{a_n^4} + 2\frac{\alpha_g^6}{a_n^6} - 5\frac{\alpha_g^8}{a_n^8} + \cdots}} \tag{7.4}$$

Compared to Sommerfeld's gravitational fine structure formula

$$E_n = \frac{E_0}{\sqrt{1 + \frac{\alpha_g^2}{a_n^2}}} \quad , \tag{7.5}$$

(7.3) is a corrected fine structure formula for the gravitational atom, which contains the Sommerfeld structure in a first approximation.

To compare the corrected fine structure formula (7.3) with the original break-off condition (7.1), we expand both formulas for $\alpha_g \ll 1$ (below, $k$ always stands for $|k|$).

For the original, implicit break-off condition (7.1), we get (also see (5.19))

$$E = E_0 \left( 1 - \frac{\alpha_g^2}{2(N+k)^2} - \frac{1}{8}\frac{4N-3k}{k(N+k)^4 k}\alpha_g^4 - \frac{1}{16}\frac{2N^3+12N^2k-18Nk^2+5k^3}{(N+k)^6 k^3}\alpha_g^6 + O(\alpha_g^8) \right). \quad (7.6)$$

For the approximate explicit break-off condition (7.3), we get

$$E = E_0 \left( 1 - \frac{\alpha_g^2}{2(N+k)^2} - \frac{1}{8}\frac{4N-3k}{k(N+k)^4 k}\alpha_g^4 - \frac{1}{16}\frac{2N^3+12N^2k-\underline{10Nk^2+13k^3}}{(N+k)^6 k^3}\alpha_g^6 + O(\alpha_g^8) \right). \quad (7.7)$$

One can see that, for $\alpha_g \ll 1$, the corrected gravitational fine structure formula (7.3) does not deviate from the original break-off condition (7.1) up to the term that is proportional to $\alpha_g^6$ (underlined term).